# Macroscopic Magneto-Chiroptical Metasurfaces

Gaia Petrucci[1], Alessio Gabbani[1,2], Esteban Pedrueza-Villalmanzo[3], Giuseppe Cucinotta[2], Matteo Atzori[4], Alexandre Dmitriev[3*], Francesco Pineider[1*]

[1]Dipartimento di Chimica e Chimica Industriale, Università di Pisa, Via Giuseppe Moruzzi 13, 56124, Pisa, Italy

[2]INSTM and Department of Chemistry "Ugo Schiff", University of Florence

[3]Department of Physics, University of Gothenburg

[4] Laboratoire National des Champs Magnétiques Intenses (LNCMI), Univ. Grenoble Alpes, INSA Toulouse, Univ. Toulouse Paul Sabatier, EMFL, CNRS, Grenoble, France

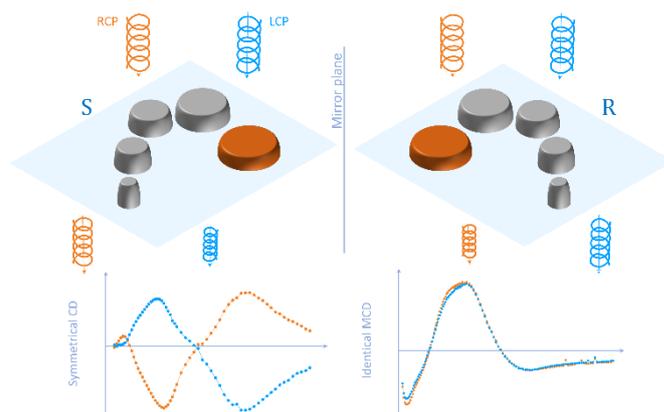

**ABSTRACT:** Nanophotonic chiral antennas exhibit orders of magnitude higher circular dichroism (CD) compared to molecular systems. Merging magnetism and structural chirality at the nanometric level allows for the efficient magnetic control of the dichroic response, bringing exciting new prospects to active nanophotonic devices and magnetochirality. Here we devise macroscale enantiomeric magnetophotonic metasurfaces of plasmon and ferromagnetic spiral antennas. Mixed 2D- and 3D-chiral nanoantennas induce large CD response, where we identify reciprocal and non-reciprocal contributions. The simultaneous chiroptical and magneto-optical response in a wide spectral range with these metasurfaces delivers an attractive platform for the study of magnetochirality at the nanoscale. Exploring further this type of magnetophotonic metasurfaces allows the realization of high-sensitivity chiral sensors and prompts the design of novel macroscopic optical devices operating with polarized light.

Chiroptical spectroscopies are traditionally used to assess the chiral nature of molecules, but recently they have raised strong interest in nanophotonics. The current rapid improvements in nanofabrication has empowered high control of the nanostructures shape and their spatial arrangement as chiral plasmon nanoantennas, allowing to achieve significantly boosted chiroptical signals compared to those of molecular systems[1,2,3]. Such significantly improved chiroptics can be exploited for chiral sensing[4], light-driven enantio-selective molecular synthesis[5], or in the development of optical modulators or light polarization control[6]. Chiral nanostructures are also potentially relevant in chiral induced spin selectivity[7]. Indeed the chirality of the nanostructures was reported to control the spin selectivity of electron transfer in CdSe-CdTe quantum dots dyads[8]. Nevertheless, the origin of chiroptical effects in three- and two-dimensional (3D, 2D) plasmon nanoantennas is currently under debate due to the inherent challenges in analyzing chiroptical response of nanoantennas with strong optical anisotropy[9,10] like those achieved with various lithographies, and the fabrication of nanostructured surfaces with perfect enantiomeric structure remains a challenging goal.

In magnetophotonics, tunability of circular dichroism by an external magnetic field is readily available by combining plasmon and magnetic elements even in structurally symmetric (achiral) nanoantennas[11,12,13,14]. When magnetic circular dichroism (MCD) and conventional optical natural circular dichroism (CD) are simultaneously present, magneto-chiral dichroism (MChD) or magnetochirality, i.e. the differential absorption of unpolarized light by a chiral magnetized medium, can be achieved by breaking both time- and space-reversal symmetries. While magnetochirality has been observed in molecular systems[15,16], its study in solid state nanostructures in still at its infancy[17,18,19].

For the coexistence of MCD and CD in metallic plasmon nanoantennas, a magnetic element should be present with the optical losses mitigated as much as possible to retain a pronounced plasmon response while the spatial symmetry of the nanoantennas needs to be broken. The latter requirement is somewhat challenging from the nanofabrication standpoint with few approaches previously employed. The symmetry breaking can be induced either at a single nanostructure or by the spatial arrangement of several distinct nanoantenna elements. In the first approach, chiral nanoparticles can be obtained using chiral templates while performing a colloidal synthesis[20], or using complex multi-step processes. For example, direct laser writing of 3D polymeric structures used as resists can be employed, followed by Au electrodeposition and polymer etching[6]. Another example is the use of physical vapor deposition to grow plasmon nano-helices under an oblique deposition angle while simultaneously rotating substrate[21]. Ramp-shaped chiral nanostructures can be also prepared by e-beam evaporation followed by focused-ion beam milling[22]. In the second approach, the 3D chiral plasmon oligomers can be created by arranging several plasmon nanodisks in a chiral geometry, using electron-beam lithography combined with layer-by-layer stacking[23]. Similar geometries can be also achieved by the macroscale bottom-up nanofabrication such as hole-mask colloidal lithography[24], with previously reported 2D optically chiral magnetoplasmonic nanoantennas showing simultaneously CD and MCD responses in the vis-near-IR spectral range; 3D-chiral antennas, where the symmetry is broken by changing the material (Ag or Au) and the thickness of the constituent nanodisk elements[25]; gold spiral-type nanoantennas can also be realized in a macroscale bottom-up chiral metasurface[26]. Here we produced enantiomeric metasurfaces with both 2D and 3D chiral nanoantennas, displaying broadband optical CD and MCD in the 500-2200 nm spectral range. Nanoantennas feature one nickel and four silver nanodisks, arranged in a spiral on the surface. The two enantiomorphs present nearly mirror-symmetric CD spectra and comparable MCD responses. We deconvolved reciprocal and non-reciprocal contributions to the chiroptical response of these metasurfaces and tested the magneto-chiral dichroism of the system.

Magneto-chiroptical metasurfaces of opposite helicity were fabricated by hole-mask colloidal lithography on cm$^2$ surfaces on glass. The nanoantennas spiral shape is formed by the Ni nanodisk with largest diameter, followed by smaller Ag nanodisks instead with larger thickness (Figure 1). Nickel was chosen as magnetic material for its stability toward oxidation and its pronounced plasmon response with reasonably low losses[27]. Figure 1 shows a sketch of the two enantiomeric magneto-chiroptical nanoantennas (hereafter defined as left-handed "S" and right-handed "R"), along with scanning electron microscopy (SEM) and atomic force microscopy (AFM) surface overview. From SEM imaging the spiral arrangement of the nanodisks is highlighted and the lateral dimensions, the density and the homogeneity of the nanostructures are evaluated. AFM elucidates the 3D character of nanoantennas elements, with nanodisks height varying from ~15 (Ni) to ~23 nm (last Ag nanodisk of the spiral). The optical extinction spectra (Figure 2a) of the two enantiomer metasurfaces are nearly superimposable.

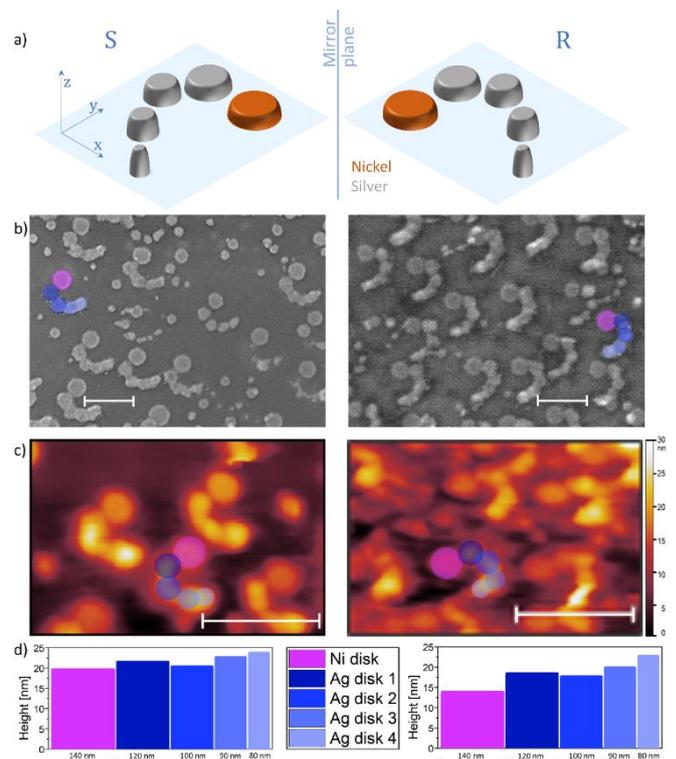

**Figure 1.** a) Sketch of the two enantiomeric nanoantennas; b) SEM images (scale bars are 200 nm) and c) AFM images (scale bars are 300 nm) of the two samples; d) height and diameter of each disk, measured on the spiral color-coded in b) and c).

As no long-range order/diffractive effects are present in these surfaces, the plasmon modes hybridization of individual nanodisks in the antenna is expected to represent the optical response of the spirals. Three resonant features are observed, centered at 500, 800 and 1500 nm respectively. While the complex arrangement of interacting nanodisks in the antenna makes the exact attribution of each resonance challenging, the extinction peak at 500 nm can be tentatively assigned to the Ag nanodisk plasmon due to its spectral range and its chiroptical response (see below). The remaining two broad and redshifted resonances can be related to the hybrid oligomer modes involving two or more nanodisks. To deconstruct the chiroptical response of these metasurfaces, a first important distinction between 2D and 3D chirality needs to be pointed out. 2D chiral nanostructures are not superimposable upon any symmetry operation while kept on the sample plane. This kind of 2D symmetry breaking leads to non-reciprocal chiroptical response, with CD reversing sign upon changing the illumination side[9,12]. This effect can be regarded as 'non-reciprocal circular dichroism', as it breaks Lorentz reciprocity, and has been observed also in polymer films containing achiral molecules[28,29,30,31]. The effect is substantially different to the circular dichroism typical of 3D chiral objects, which on the other hand is a reciprocal effect, i.e. it gives a CD signal that does not change sign upon sample/illumination side flip. Both effects are, however, invariant to in-plane rotation, making them substantially different from linear dichroism (LD) that is strongly affected by in-plane structure rotation. The plane invariance and illumination-side reciprocity were exploited to deconvolute the chiroptical effects in these metasurfaces and separate them from the linear dichroism, also present here.

In general, non-reciprocal CD in nanoantennas arises from in-plane broken rotational symmetry[9]. In order to earn a reciprocal CD response, the symmetry needs to be broken also out-of-plane (along the z direction reported in Figure 1). In the case of spiral nanoantennas, both cases are realized. Reciprocal CD (hereafter CD) arises from the different heights of the nanodisks, and the fact that the nanostructures are supported on a glass surface: both factors break the symmetry along z. Non-reciprocal CD, on the other hand, stems from the planar spiral arrangement of the nanodisks which is responsible for the in-plane rotational symmetry breaking. Both contributions relate to the excitation of plasmon resonances that are the hybrids of the individual nanodisks plasmons.

Figures 2b,c report the CD and non-reciprocal CD spectra for left- and right-handed enantiomeric metasurfaces, with the dissymmetry factor $g$ calculated for the CD as the ratio between the CD and the extinction,

$$g = 2\frac{A_{LCP}-A_{RCP}}{A_{LCP}+A_{RCP}} = \frac{\Delta A}{A}, \quad (1)$$

where $A_{LCP}$ and $A_{RCP}$ denote the optical extinction of left and right circular polarized light. It should be noticed that for wavelengths greater than 1800 nm the $g$ factor is not reliable due to vanishing value of the optical extinction. The CD was collected following a procedure adapted from Harada et al.[10], which allows measuring simultaneously CD and LD, and canceling out LD-related artifacts in the CD signal. First, the in-plane rotation angle at which the linear dichroism (LD) signal is maximized is found (see Figures S2, SI). The angle corresponding to the LD maximum is identified as the 0° of the sample, defining in this way its optical axis. The orientation of the optical axis is detected for the peak at 860 nm, and for the one at 1400 nm. Distinct optical axes were found for the two maxima, with the zero position at 1400 nm differing by ~15° from the one at 860 nm, suggesting that the two peaks are related to excitations of coupled nanodisks in the antenna occurring along different directions. In both cases, the LD maximum is found to reverse sign upon 90° in-plane rotation. It follows that by acquiring the total CD along perpendicular directions and then averaging them, the LD artifacts in the CD signal are canceled out, giving almost identical results for the two couples of orthogonal angles investigated (+45/-45° and 0/+90°, see Supporting Information). We report here the CD measurements at ±45° from the zero position, as at these orientations the value of the LD signal approaches zero, thus minimizing the LD-related artifacts (Equation 2).

$$Total\ CD(f/b) = \frac{[CD(+45°)+CD(-45°)]_{f/b}}{2}, \quad (2)$$

The CD is then isolated from the non-reciprocal CD by averaging the total CD signal measured with front and back (f, b) light incidence (Equation 3), which are defined according to SI. Finally, subtracting this from the total CD, the non-reciprocal CD is recovered as well (Equation 4).

$$CD = \frac{CD_f+CD_b}{2}, \quad (3)$$

$$Non-reciprocal\ CD = Total\ CD(f/b) - CD. \quad (4)$$

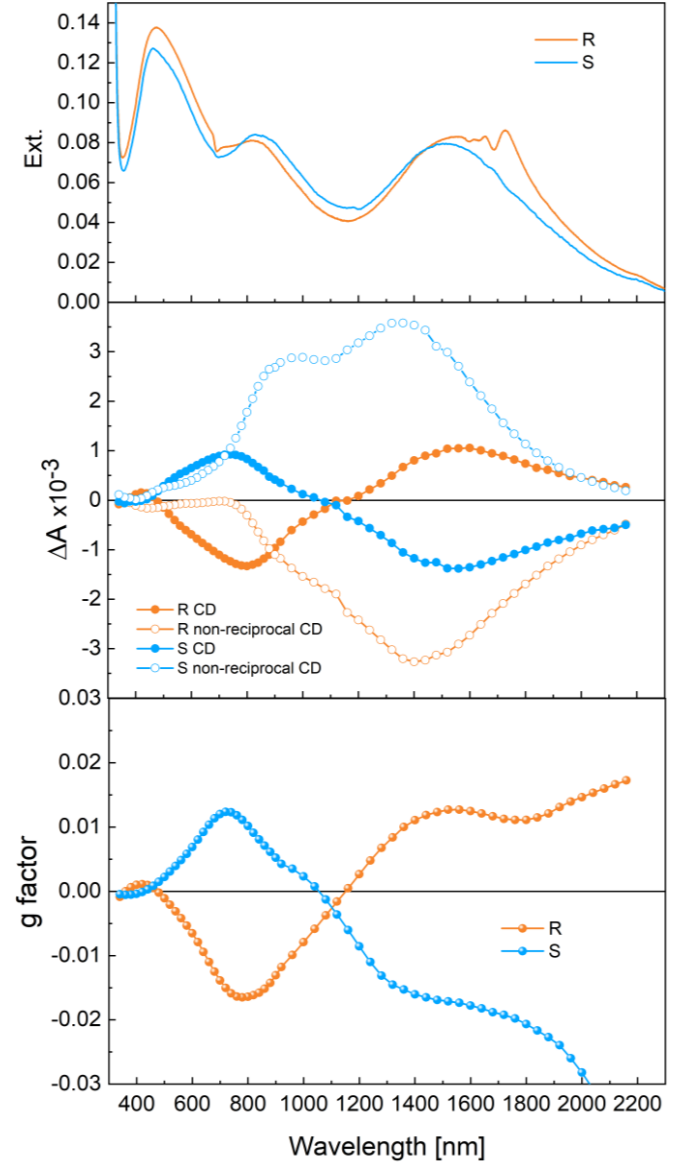

**Figure 2.** a) Extinction spectra; b) CD spectra; and c) g factor of the magneto-chiroptical metasurfaces. The plotted non-reciprocal CD is the one acquired from the front, the one from the back is reverted in sign (not reported here for clarity; see Supporting Information).

As expected for two enantiomers, the CD contribution in left- and right-handed metasurfaces is of the same shape but inverted in sign (Figure 2c). The shape of the non-reciprocal CD contribution slightly differs for the two enantiomers, but the signal acquired from the same side of each sample has similarly inverted sign. It is worth mentioning that for these metasurfaces the contribution of non-reciprocal CD is higher than the one of reciprocal CD, highlighting the importance of the in-plane rotational symmetry breaking induced by the 2D spiral arrangement of the nanodisks in the nanoantennas.

MCD spectra (Figure 3) were acquired using the same experimental setup used for CD but placing the samples in a split coil electromagnet able to apply a magnetic field up to 1.4 T along the light propagation direction. Similarly to optical extinction, the MCD signal of left- and right-handed enantiomeric metasurfaces is superimposable. Three spectral features are visible: a negative dip at 370 nm, a broader intense peak at 900 nm, and a weaker and broad negative peak 1400 nm. MCD hysteresis loops (inset in Figure 3) were acquired at fixed wavelength by scanning the magnetic field between ±1.4 T. The MCD signal saturates at 0.35 Tesla for both the enantiomers, consistently with the magneto-optical behavior of nickel, which is dominant at all the inspected wavelengths (Figure S4, SI), suggesting full magneto-optical coupling between silver and nickel nanodisks in the antenna [32].

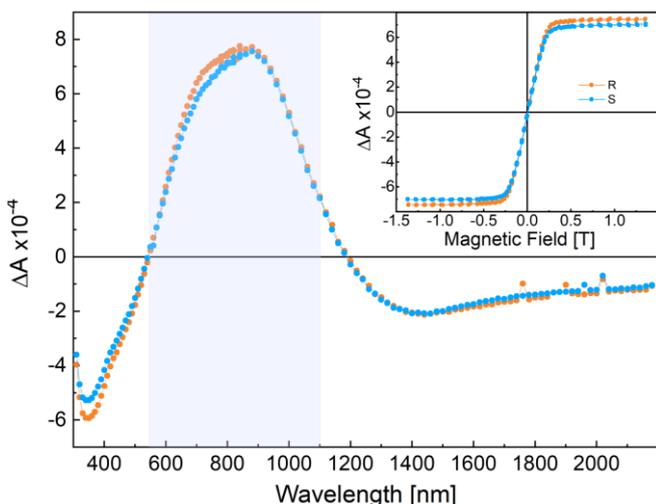

**Figure 3.** MCD of the magneto-chiroptical metasurface enantiomers. In the inset the hysteresis loops collected at the wavelength of the MCD maximum are reported. The region of maximum overlap between CD an MCD signals is highlighted in light blue.

Preliminary Magneto-chiral dichroism (MChD) measurements were performed by unpolarized light illumination under an applied magnetic field along the light propagation direction. In principle, the effect is revealed by detecting a dichroic signal that changes sign either by changing the enantiomer or inverting the relative direction of the applied magnetic field and light wavevector. Magnetochirality is expected in a spectral range where CD and MCD signals overlap[16] (see highlighted region in Figure 3). The measurements were performed at different temperatures using a variable temperature insert for low temperature measurements (down to 4 K) (see SI for details), applying magnetic field of various magnitude (within ±2.0 T) and changing the relative orientation of magnetic field and incident light. However, it was not possible to detect a MChD signature in these metasurfaces, likely due to the overall weak optical extinction that leads to MChD signal below the detection limit of our setup, which was of the order of $10^{-4}$. A common rule of thumb to estimate the order of magnitude of the MChD signal is to calculate it as the product of CD and MCD. In this assumption, the MChD expected for these metasurfaces in the highlighted region is of the order of $10^{-7}$ [33,3].

Summarizing, macroscopic magneto-chiroptical metasurface enantiomers were realized by the bottom-up nanofabrication on several $cm^2$. The two enantiomers gave superimposable optical extinction and MCD spectra, with the CD and dissymmetry factor g disentangled to separate contributions due to 2D and 3D chirality of the nanoantennas building the metasurfaces. MCD measurements highlighted the strong interaction of magnetic (Ni) and plasmonic (Ag) nanodisks in the spiral antennas, the latter acting as highly magneto-optically coupled systems. Increasing both CD and MCD of such metasurfaces would potentially lead to a very pronounced magneto-chiral dichroism.

**Acknowledgements**
Sample fabrication was performed at Myfab, Chalmers. F.P., G.P. and A.G. acknowledge for funding PRIN2017 Grant No. 2017CR5WCH Q-ChiSS (Italian MIUR) and PRA_2017_25 (Università di Pisa). Professor Gennaro Pescitelli is acknowledged for the fruitful scientific discussion.

**DATA AVAILABILITY**
The data that support the findings of this study are available within the article and its supporting information.

**AUTHOR INFORMATION**
*E-mail: (F.P.) francesco.pineider@unipi.it
*E-mail: (A.D.) alexd@physics.gu.se

# Macroscopic magneto-chiroptical metasurfaces

Gaia Petrucci[1], Alessio Gabbani[1,2], Esteban Pedrueza-Villalmanzo[3], Giuseppe Cucinotta[2], Matteo Atzori[4], Alexandre Dmitriev[3*], Francesco Pineider[1*]

[1]Dipartimento di Chimica e Chimica Industriale, Università di Pisa, Via Giuseppe Moruzzi 13, 56124, Pisa, Italy

[2]INSTM and Department of Chemistry "Ugo Schiff", University of Florence

[3]Department of Physics, University of Gothenburg

[4] Laboratoire National des Champs Magnétiques Intenses (LNCMI), Univ. Grenoble Alpes, INSA Toulouse, Univ. Toulouse Paul Sabatier, EMFL, CNRS, Grenoble, France

*francesco.pineider@unipi.it

*alexd@physics.gu.se

## Supporting Information

**Nanofabrication - Hole-mask colloidal lithography (HCL)**

A polymethyl methacrylate film (PMMA 495 4A from Michrochem, refractive index of 1.491 at 500 nm) is spin-coated on the glass substrate (3000 rpm for 1 min, nominal thickness 250 nm), plasma etched with oxygen for 5 seconds to improve hydrophilicity (PlasmaTherm Reactive Ion Etcher at 50 W) and then positively charged using an electrolytic solution of poly(diallyl dimethyl ammonium) chloride (PDDA). A colloidal solution of polystyrene (PS) beads (Aldrich), which possess negative surface charge, is then drop-casted on it. Within a few minutes the beads land on the substrate creating a disordered array of evenly distributed spheres. The exceeding solution is then removed with a smooth and regular ultra-pure water flow, and then dried with an intense nitrogen flow, to prevent the landing beads from rearranging. A sacrificial layer of 100 nm of chromium is evaporated on the PMMA and on the beads using an E-beam thin film evaporator PVD - 225 from Lesker, equipped with a 12 pockets electron gun for 12 sources and with a cryo-pump that gives a base pressure of $4 \cdot 10^{-8}$ mbar. The beads are then removed by tape-stripping, leaving nanoholes in the chromium film, and by a more intense oxygen reactive ion etching (9 min), the parts of the polymer that are uncovered by the metallic film are selectively removed, generating the mask. The metal (Ni, Ag) is then evaporated through the holes while imposing a defined tilt and rotation position to the sample holder, drawing arrays of well-defined short-range-ordered dotted nanospirals. Finally, the mask is removed by lift-off in acetone.

**Characterizations**

SEM images were acquired using a Zeiss Supra 60 VP equipped with a vacuum Gemini column.

AFM measurements were performed with a NT-MDT P47-PRO instrument (NT-MDT, Zelenograd Russia).

Extinction spectra were collected on a Jasco V-670 spectrophotometer operating in the range between 190 and 2700 nm.

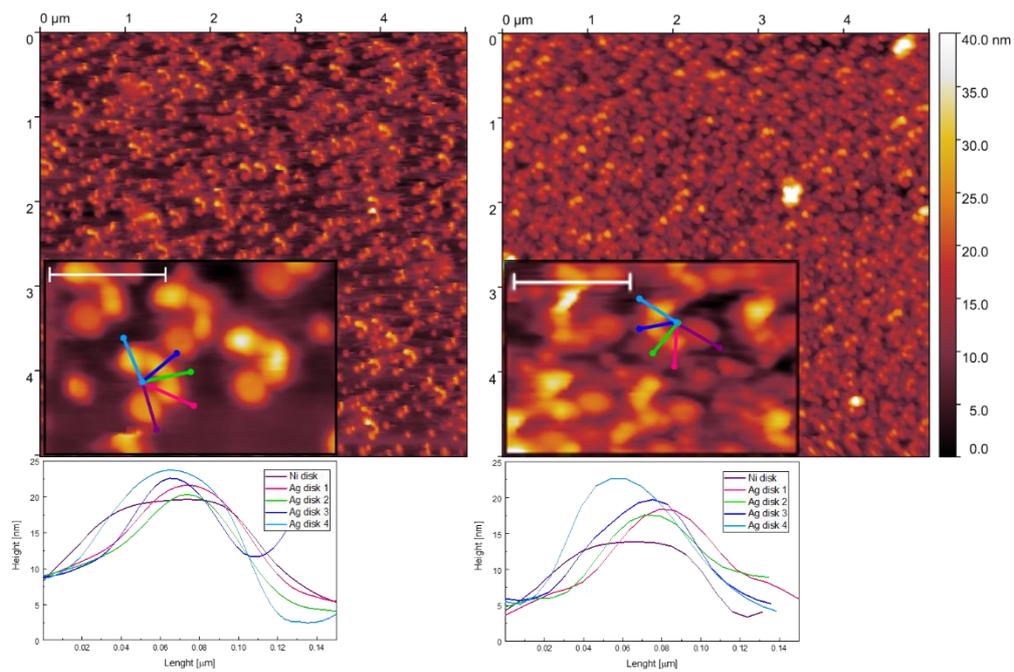

*Figure S1 Wide areas AFM images of left- and right-handed magneto-chiroptical metasurfaces. Maximum height in the insets is 30 nm and scale bars are 300 nm. The highlighted profiles are color-coded and reported below each image. The left column refers to sample S, while the right one to R.*

## Circular dichroism and magnetic circular dichroism measurements (CD-MCD)

The CD spectra of the magneto-chiroptical metasurfaces were acquired at room temperature in the spectral range between 300 nm and 2200 nm using a home-built setup based on polarization modulation technique and capable of simultaneously acquiring circular dichroism (CD) and linear dichroism (LD) signals. Samples were mounted on a rotating stage to measure CD at different in-plane rotation angle, while maintaining light incidence normal to the sample substrate.

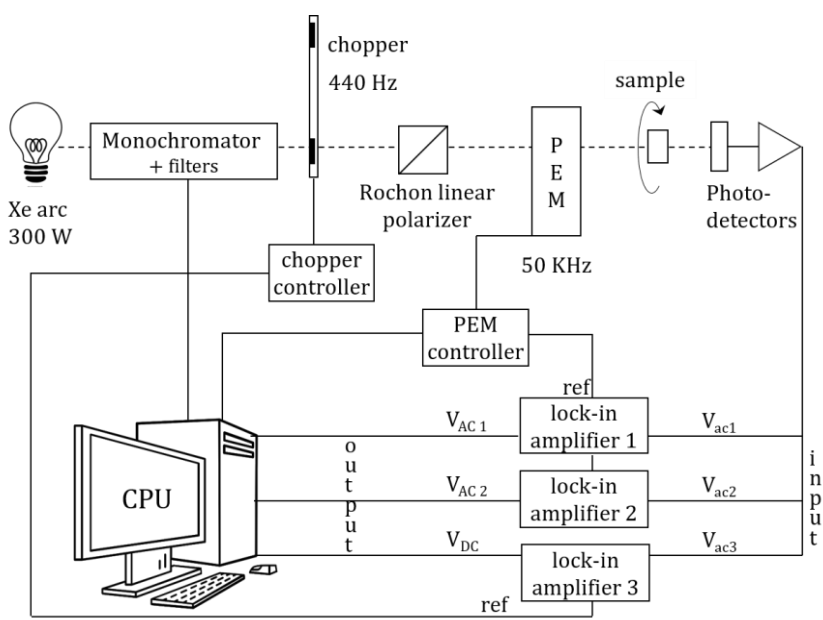

*Figure S2. Block diagram of the CD setup. Lenses are not reported.*

The setup used to collect CD (sketched in Figure S2) is equipped with a tunable light source (TLS260, Newport) comprising a Xe lamp continuum light source working in the 200-2500 nm range (power 300 W), a filter wheel and a monochromator (Oriel Cornerstone 260). Light is modulated at 440 Hz through an optical chopper (MC1000A, Thorlabs) to filter out ambient noise. The output light from the monochromator is linearly polarized with a Rochon prism oriented at 45°, after which a photo-elastic modulator (PEM100, Hinds Instruments) placed before the sample further modulates the polarization at a frequency of 50 kHz. PEM retardation is set to 0.383 λ in order to detect simultaneously CD and LD at the first and second harmonic of the PEM frequency respectively [1]. Two detectors are alternatively used to collect the light transmitted by the sample, a photomultiplier tube for the UV-visible range (300-850 nm), and an InGaAs photodetector for the near-infrared (750-2500 nm). Three lock-in amplifiers are used, two of them locked at the first and second harmonics of the PEM (SR850 and SR830, Stanford Research Systems), in order to collect at the same time CD and LD signals, that are the main contributions found at the two frequencies respectively; and a third one locked at a chopper frequency (SR7280, Ametech Advanced Measurements Technology), acquiring the total signal. The sample holder is provided with a goniometer for in-plane rotation of the sample, and is designed to allow mounting the sample with light illumination from the front and back sides, which are defined as the ones exposing the air interface and the glass substrate respectively. To ensure normal incidence of light on the sample, light reflected from the sample is checked. For MCD measurements the setup was modified placing the samples inside a split-coil electromagnet (45 mm, 3470 GMW) operating in the range ± 1.4 T, and only the signal at the first harmonic of the PEM is used (in this measurement PEM retardation is set to 0.25 $\lambda$). Lenses are used to focalize light on the sample and on the detectors. For both cases the dichroic signal is recovered as:

$$\Delta A = C \frac{V_{AC}}{V_{DC}}$$

with $\Delta A$ being the difference in absorption of the two light polarizations investigated, $\Delta A = A_{LCP} - A_{RCP}$, $V_{AC}$ the signal modulated at the PEM frequency, $V_{DC}$ the one at the chopper frequency and $C$ is a calibration factor evaluated using a standard solution of $K_3Fe(CN)_6$ [2].

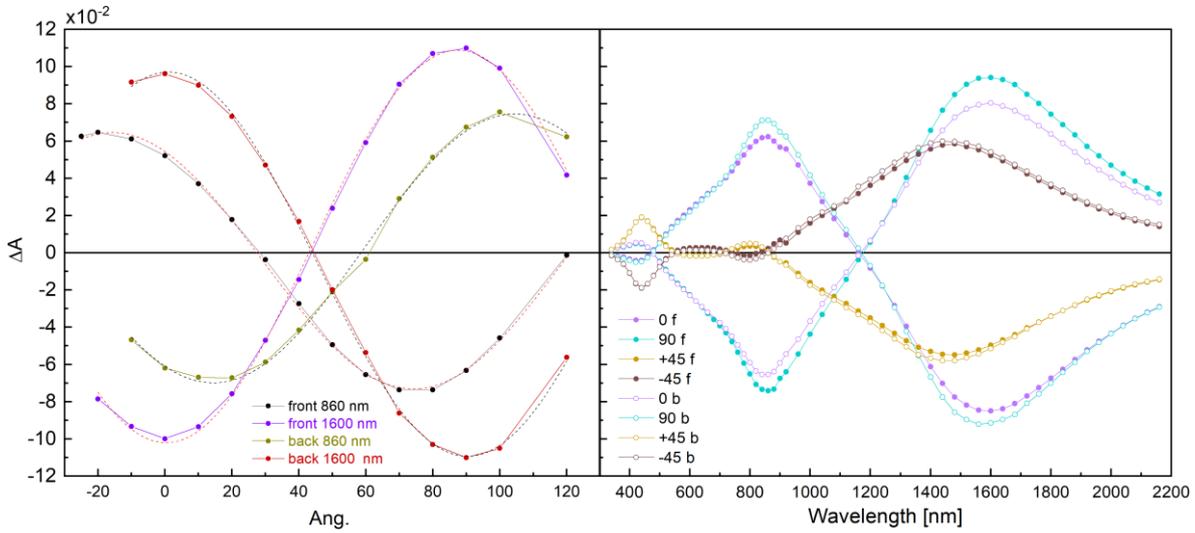

*Figure S3. Left: angular plot of the LD signal at the wavelengths of the two LD maxima from front and back light illumination for left-handed metasurface; each plot is fitted with a sine function; Right: LD spectra of left-handed magneto-chiroptical metasurface enantiomer.*

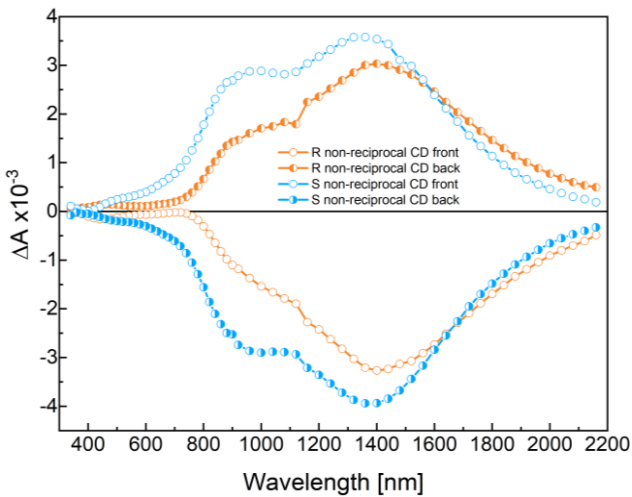

*Figure S4. Front and back non-reciprocal CD of left- and right-handed magneto-chiroptical metasurface enantiomers.*

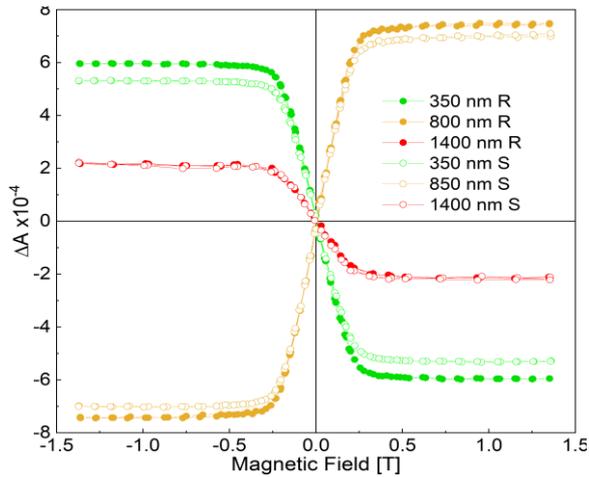

*Figure S5. MCD hysteresis loops of left- and right-handed magneto-chiroptical metasurface enantiomers acquired a different wavelengths.*

**Magneto-Chiral Dichroism measurements**

MChD experiments were performed with a home-made multichannel MChD spectrometer operating in the visible and near infrared spectral window (420–1600 nm) between 4.0 and 300 K with an alternating magnetic field ***B*** up to ±2 T. A detailed description of the measurement apparatus has been reported elsewhere [3]. MChD spectra were acquired on left- and right handed metasurfaces. The samples were mounted on a titanium sample holder and irradiated with unpolarized light over a 1.0 mm diameter hole centered with respect to a 1.0 mm diameter collimated beam. Measurements were performed in the 4.0−150 K range with an alternating magnetic field ***B*** = ±1.0 T and frequency $\Omega$ = 0.04 Hz. MChD spectra as a function of the magnetic field were recorded at *T* = 4.0 K for alternating magnetic fields of different amplitudes (0.0-1.91 T). The MChD spectra were obtained at each temperature/field by recording, on average, 50.000 spectra with an integration time, $t_{int}$ of 7 ms, every 10 ms. Unpolarized light was provided by Thorlabs broadband Tungsten-Halogen (Thorlabs SLS201L) or Light Emitting Diodes (Thorlabs MCWHF2) light sources. The data were collected with a OPTOSKY detector equipped with a thermoelectric cooled sensor operating in the 420–1000 nm spectral region. Each spectrum was correlated to a specific magnetic field value by a dual channel digitizer (Picoscope 5000B) acquiring simultaneously triggers from the spectrometer and the magnetic field from a Hall effect sensor placed in proximity of the sample. Data are then post-processed as a synchronous detection with a MatLab routine to obtain MChD spectra.

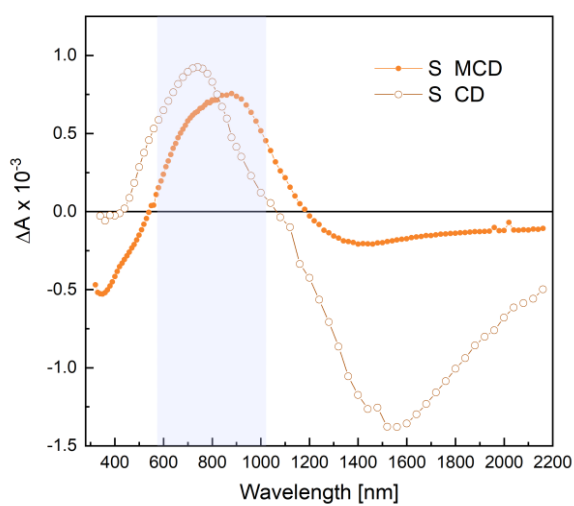

*Figure S6. CD and MCD of the spiral nanoantennas. The light blue rectangle highlights the overlapping spectral region of CD and MCD which has been used to probe MChD.*